\documentclass[12pt]{article}
\usepackage{amsfonts}
\usepackage{amsmath}
\usepackage{amssymb}
\usepackage{amscd}
\usepackage[dvips]{graphicx}

\begin{document}

\begin{center}
{\Large \textbf{Braided Tensor Products and the Covariance of Quantum
Noncommutative Free Fields}}
\end{center}

\bigskip

\begin{center}
{{\large ${\mathrm{Jerzy\;Lukierski}}$, ${\mathrm{Mariusz\;Woronowicz}}$ }}

\bigskip

{
$\mathrm{~Institute\;of\;Theoretical\;Physics}$}

{
$\mathrm{\ University\; of\; Wroclaw\; pl.\; Maxa\; Borna\; 9,\; 50-206\;
Wroclaw,\; Poland}$}

{
$\mathrm{\ e-mail:\;lukier@ift.uni.wroc.pl;woronow@ift.uni.wroc.pl}$}

\bigskip
\end{center}
\begin{abstract}
We introduce the free quantum noncommutative fields as described by braided
tensor product. The multiplication of such fields is decomposed into three
operations, describing the multiplication in the algebra $\mathcal{M}$ of
functions on noncommutative space-time, the product in the algebra $\mathcal{%
H}$ of deformed field oscillators, and the braiding by factor $\Psi _{%
\mathcal{M},\mathcal{H}}$ between algebras $\mathcal{M}$ and $\mathcal{H}$.
For noncommutativity of quantum space-time generated by the twist factor we
shall employ the $\star $-product realizations of the algebra $\mathcal{M}$
in terms of functions on standard Minkowski space. The covariance of single
noncommutative quantum fields under deformed Poincare symmetries is
described by the algebraic covariance conditions which are equivalent to the
deformation of generalized Heisenberg equations on Poincare group manifold.
We shall calculate the braided field commutator covariant under deformed
Poincare symmetries, which for free quantum noncommutative fields provides
the field quantization condition and is given by standard Pauli-Jordan
function. For ilustration of our new scheme we present explicit calculations
for the well-known case in the literature of canonically deformed free
quantum fields.
\end{abstract}

\section{Introduction}

\bigskip

It is at present a common view that the most important problem in the theory
of fundamental interactions is to quantize in consistent way the dynamical
theory of space-time, i.e. gravity, and incorporate it into the description
of all other interactions. The consistent quantization of gravitation theory
is searched in different ways: by employing new (nonlocal, discrete)
descriptions of Einstein gravity (loop quantum gravity \ \cite{1}, spin foam
models \cite{2}) or by embedding standard gravity in larger dynamical
framework ((super)-string \cite{3}, brane world models \cite{4}). In most of
these approaches the classical commutative space-time geometry is modified
because the quantization of gravitational dynamics leads to new algebraic
structure of relativistic phase space with the noncommutative quantum
space-time sector. In particular if we consider the quantum fields used e.g
in the formulation of Standard Model in the presence of quantum gravity
background, the appearance of quantum space-time implies that the standard
framework for classical and quantum fields should be suitably deformed.

In recent fifteen years there were proposed various ways of introducing the
deformation of quantum free fields which due to quantization procedure of
gravity (see e.g. \cite{lit1}) or quantized $D=10$ string effects in
tensorial gauge fields background (see e.g. \cite{lit4}) were defined on the
noncommutative space-time \cite{uog}-\cite{bala}. \ It was proposed that the
fields defined on relativistic space-time (Minkowski space) should be
replaced by new fields which algebraically take into account the quantum
gravity effects and are defined on quantum space-time. In these new
field-theoretic models one replaces the commutative space-time coordinates $%
x_{\mu }$ by the noncommutative ones%
\begin{equation}
\lbrack x_{\mu }\,,x_{\nu }]=0\qquad \Longrightarrow \qquad \lbrack \hat{x}%
_{\mu }\,,\hat{x}_{\nu }]=\frac{i}{\kappa ^{2}}\Theta _{\mu \nu }(\kappa 
\hat{x})\,,  \label{nnn}
\end{equation}%
where ($\kappa $- masslike deformation parameter) $\Theta _{\mu \nu }(\kappa 
\hat{x})=\Theta _{\mu \nu }^{(0)}+\kappa \Theta _{\mu \nu }^{(1)\rho }%
\widehat{x}_{\rho }+\cdots \,.$The choice $\Theta _{\mu \nu }(\kappa 
\widehat{x})=\Theta _{\mu \nu }^{(0)}$ \ corresponds to the so-called
canonical deformation \cite{lit1}, \cite{lit4} and $\Theta _{\mu \nu
}(\kappa \widehat{x})=\Theta _{\mu \nu }^{(1)\rho }\widehat{x}_{\rho }$
provides the Lie-algebraic noncommutativity of space-time with the most
known special case describing the $\kappa $-deformation of Minkowski space 
\cite{zak}-\cite{7}. It has been also shown in the examples of deformed
quantum free fields that the noncommutativity of space-time algebra is
linked in various ways with the deformation of field oscillators algebra
(see e.g. \cite{16}-\cite{fi}).

One introduces the symmetries in theories with noncommutativity (\ref{nnn})
in two distinct ways:

\begin{enumerate}
\item One keeps the classical Poincare symmetries and the noncommutativity (%
\ref{nnn}) is treated as introducing the terms which break the classical
relativistic symmetries \cite{lit1},\cite{fil}-\cite{pia2}. In such a way
the choice of function $\Theta _{\mu \nu }(\kappa \hat{x})\,$ can be
arbitrary (modulo Jacobi identities) and the set of\ constant tensors $%
\Theta _{\mu \nu }^{\quad \rho _{1}\ldots \rho _{n}}(\rho =0,1,2\ldots )$
describe the classical Poincare symmetry breaking parameters.

\item One can introduce the modified quantum Poincare-Hopf symmetries in a
way which leads to the covariance of the relations (\ref{nnn}) (see e.g. 
\cite{lit8}-\cite{bala}). In such approach quantum covariance means that the
relation (\ref{nnn}) has the same algebraic form in all deformed Poincare
frames and $\Theta _{\mu \nu }^{(n)\rho _{1}\ldots \rho _{n}}(n=0,1,2\ldots
) $ can be treated as set of constant parameters, which also enter into the
deformed Poincare-Hopf algebra. It appears that such method is more
restrictive, it selects only the particular classes of functions $\Theta
_{\mu \nu }(\kappa \hat{x})\,$occuring in (\ref{nnn}).

In this paper we shall follow the second approach.
\end{enumerate}

The basic aim of this paper is to study new framework for description of
covariant twist-deformed free quantum fields, with the covariant deformation
of field oscillators algebra. It is known already from earlier studies of
Faddeev-Zamolodchikov and $q$-deformed Heisenberg algebras (see e.g. \cite%
{ab}-\cite{fiore}) that the deformed commutators of field oscillators which
are covariant under the quasitriangular quantum symmetry takes a braided
form, with the covariant braid factor described by corresponding universal $%
\mathcal{R}$-matrix (see e.g. \cite{sm}, Sect. 7.2). Such $\mathcal{R}$%
-matrix-dependent braid factor will also characterize the c-number
commutator of deformed free quantum fields as well as the braided products
of these fields.

In Sect. 2 we describe the properties of general twisting procedure of Hopf
algebras and specify the description for the canonical twist. In Sect. 3 we
shall consider the deformed noncommutative free quantum fields $\phi (%
\widehat{x})$ as described by binary braided tensor products $\mathcal{M}%
\otimes \mathcal{H}$ where $\mathcal{M}$ is the algebra of functions on
noncommutative space-time and $\mathcal{H}$ denotes the algebra of deformed
field oscillators (see also \cite{mou}-\cite{pic}, where however the
unbraided tensor product is used). We point out that the nontrivial braiding
between quantum symmetry modules $\mathcal{M}$ and $\mathcal{H}$ will be
only revealed in the multiplication procedure of quantum fields. In Sect. 4
we describe two covariant actions of the deformed Poincare generators on a
deformed quantum field. In the definition of single noncommutative quantum
field after introduction in $\mathcal{M}$ of the Weyl map the tensor
structure $\mathcal{M}\otimes \mathcal{H}$ collapses into the field
oscillators algebra module $\mathcal{H}$, what is also required by the
correct \ no-deformation limit of the deformed quantum fields. Algebraically
the covariance condition will be expressed on deformed fields as generalized
Heisenberg field equations. Further in Sect. 5 we shall consider the
products of deformed free fields with the braided multiplication $m_{\bullet
}$. Such new multiplication rule introduces the braiding factor $\Psi _{%
\mathcal{M},\mathcal{H}}$ between the algebra of space-time functions and
deformed field oscillators (see also Fiore and Wess \cite{wess},\cite{fi},%
\cite{fiore}\footnote{%
Similar braiding factor has been considered also in \cite{tur} but for
different aims.}). If the deformation of Poincare symmetries is described by
universal $\mathcal{R}$-matrix (we denote $\mathcal{R=R}_{(1)}\otimes 
\mathcal{R}_{(2)}$ and $\mathcal{R}_{21}\mathcal{=R}_{(2)}\otimes \mathcal{R}%
_{(1)}$) the braid factor\ $\Psi _{\mathcal{M},\mathcal{H}}$ will be defined
by $\mathcal{R}_{21}$ and the quantization of free quantum field ${\phi }\in 
\mathcal{M}\otimes \mathcal{H}$ will be described by the covariant braided
commutator\footnote{%
We introduced the short-hand notation. The operator $\mathcal{R}_{21}$ acts
in the definition (2) of braided commutator as follows
\par
\begin{center}
$\mathcal{R}_{21}\blacktriangleright (\phi (\widehat{y})\bullet \phi (%
\widehat{x}))\equiv m_{\bullet }\circ \mathcal{R}_{21}\blacktriangleright
(\phi (\widehat{y})\otimes \phi (\widehat{x}))=(\mathcal{R}%
_{(2)}\blacktriangleright \phi (\widehat{y}))\bullet (\mathcal{R}%
_{(1)}\blacktriangleright \phi (\widehat{x})).$%
\end{center}
\par
{}} 
\begin{equation}
\lbrack \phi (\widehat{x}),\phi (\widehat{y})]_{\bullet }^{BR}\equiv \phi (%
\widehat{x})\bullet \phi (\widehat{y})-\mathcal{R}_{21}\blacktriangleright
(\phi (\widehat{y})\bullet \phi (\widehat{x})),  \label{fgf}
\end{equation}%
where we shall argue in Sect. 4 that the action $g\blacktriangleright \phi $
of deformed Poincare algebra generators occurring in $\mathcal{R}_{21}$ is
fully described in covariant theory by the operation $\blacktriangleright $
of quantum adjoint in the algebra $\mathcal{H}$ of deformed field
oscillators. Our basic point is here (see also Aschieri at all in \cite{alv}%
) that the deformed Poincare covariance of the free quantum field commutator
requires the braided form (\ref{fgf}).

In the paper we restrict our explicit considerations to the triangular
deformations described by Drinfeld twist \cite{drinfeld}. Our main aim is to
show that using suitable twist-covariant braided algebra of deformed field
oscillators we get the c-number braided field commutator (\ref{fgf})
describing covariant algebra of quantum noncommutative free fields. Because
twist deformation of Poincare symmetries does not modify the mass-shell
condition, such braided commutator of twist-deformed quantum free fields is
given by the standard relativistic Pauli-Jordan commutator function.

In order to illustrate our approach we shall consider in explicit way the
simple example of canonically deformed QFT ( $\Theta _{\mu \nu }(\kappa 
\widehat{x})=\Theta _{\mu \nu }^{(0)}$ in (\ref{nnn})), with Weyl map
described by Moyal-Weyl $\star $-product. It can be also shown (see e.g. 
\cite{lw}), that important class of noncanonically deformed relations (\ref%
{nnn}) with rhs containing the linear and quadratic terms can be made
covariant under twisted Poincare symmetries. On such class of noncommutative
space-times one can define as well the corresponding deformed free quantum
fields with covariant braided field commutator (\ref{fgf}). We conclude that
our scheme provides the description of deformed free QFT which is covariant
under twisted Hopf-Poincare algebra but in principle can be extended to the
deformations described by any quasitriangular universal $\mathcal{R}$-matrix.

\section{The quantum Poincare-Hopf algebras and the deformation by canonical
twist}

The classical relativistic symmetries are generated by Poincare Lie algebra $%
\mathcal{P}$, which can be endowed with trivial (primitive) coalgebraic
structure by the coproducts $(g=(M_{\mu \nu },P_{\mu }))$%
\begin{equation}
\Delta _{0}(g)=g\otimes 1+1\otimes g,
\end{equation}%
describing the homomorphic map $\Delta _{0}:\mathcal{U}_{0}(\mathcal{P})%
\mathcal{\rightarrow \mathcal{U}}_{0}\mathcal{(\mathcal{P})\otimes U}_{0}(%
\mathcal{P})$ of enveloping classical Poincare algebra $\mathcal{U}_{0}(%
\mathcal{P})$. For the description of quantum relativistic symmetries one
uses the quantum Poincare-Hopf algebras $\mathbb{H}(\mathcal{U}(\mathcal{P}%
),m,\eta ,\Delta ,\epsilon ,S),$ where $\mathcal{U}(\mathcal{P})$ describes
the enveloping deformed Poincare algebra with multiplication map $m:\mathcal{%
U}(\mathcal{P})\otimes \mathcal{U}(\mathcal{P})\rightarrow \mathcal{U}(%
\mathcal{P})$; $\Delta :\mathcal{U}(\mathcal{P})\mathcal{\rightarrow 
\mathcal{U}(\mathcal{P})\otimes U}(\mathcal{P})$ denotes the nonprimitive
(noncommutative) coproduct map and $S:\mathcal{U}(\mathcal{P})\mathcal{%
\rightarrow \mathcal{U}(\mathcal{P})}$ defines the antipode or coinverse,
which for classical Lie algebras is given by $S_{0}(g)=-g$. \ For complete
definition of a Hopf algebra (see e.g. \cite{maj}) one introduces also the
unit $\eta $ ($m(\eta \otimes a)=a$) and counit $\epsilon $, playing the
role of unit for coalgebra maps.

A simple way of passing from classical to quantum Poincer-Hopf algebra is
achieves by introducing the twist $\mathcal{F}$. During twisting procedure
of undeformed Poincare Hopf algebra the algebraic sector remains unchanged,
therefore the classical Poincare algebra remains valid and the formulae for
Casimirs are not changed. The twist factor $\mathcal{F}\in \mathcal{U}(%
\mathcal{P})\otimes \mathcal{U}(\mathcal{P})$ deforms only the co-structure
of Hopf algebra by modification of coproduct and antipode by the following
formulae 
\begin{align}
& \Delta ^{\mathcal{F}}(g)=\mathcal{F}\Delta _{0}(g)\mathcal{F}^{-1}\,,
\label{ppoc} \\
& S^{\mathcal{F}}=vS_{0}(g)v^{-1}=-vgv^{-1}\,,\qquad v=f_{(1)}S(f_{(2)})\,,
\end{align}%
where we denote $\mathcal{F}=f_{(1)}\otimes f_{(2)}\footnote{%
We use Sweedler notation with suppresing the summ index.}$. When the
coproduct is changed the action of algebra $\mathcal{U}(\mathcal{P})$ on
tensor product representation is modified \cite{maj} because the coproduct (%
\ref{ppoc}) enters into the covariant action as follows $(\Delta ^{\mathcal{F%
}}(g)=g_{(1)}\otimes g_{(2)})$ 
\begin{equation}
g\triangleright (vw)=m[\Delta ^{\mathcal{F}}(g)\triangleright (v\otimes
w)]=\sum (g_{(1)}\triangleright v)(g_{(2)}\triangleright w)\,,\qquad
g\triangleright 1=\epsilon (g)\triangleright 1\,.  \label{vvoc}
\end{equation}

If twist $\mathcal{F}$ does not generate nontrivial coassociator it should
satisfy the 2-cocycle condition

\begin{equation}
\mathcal{F}_{12}(\Delta \otimes id)\mathcal{F}=\mathcal{F}_{23}(id\otimes
\Delta )\mathcal{F}.
\end{equation}%
In such a case the leading term describing twist factor $\mathcal{F}$ is
described by the classical $r$-matrix (by $\xi $ we denote the deformation
parameter)

\begin{equation}
\mathcal{F}=1\otimes 1+\xi r+\mathcal{O(\xi }^{2}\mathcal{)},
\end{equation}%
where $r\in g\wedge g$ satisfies classical Yang-Baxter equation (CYBE), one
gets for any generator $g$ that

\begin{equation}
\Delta \mathcal{(}g\mathcal{)}=\Delta ^{(0)}\mathcal{(}g\mathcal{)+}\xi
\lbrack r,\Delta ^{(0)}\mathcal{(}g\mathcal{)}]+\mathcal{O}(\xi ^{2}).
\label{exx}
\end{equation}%
From Lie algebra relations for $g$ follows that the first order correction
in (\ref{exx}) belongs to the tensor product $g\otimes g$.

The algebra of the canonically deformed space-time 
\begin{equation}
\lbrack \hat{x}_{\mu }\,,\hat{x}_{\nu }]=i\theta _{\mu \nu }\,,  \label{kk}
\end{equation}%
can be described as covariant under deformed (twisted) Poincare symmetries
if one introduces the canonical twist factor \cite{dlw6},\cite{lit8},\cite%
{lit8a} 
\begin{equation}
\mathcal{F}=e^{\frac{\,i}{2}\,\theta ^{\mu \nu }\,P_{\mu }\otimes P_{\nu
}}\,\,.  \label{twist0}
\end{equation}

For canonical deformation, from twist factor (\ref{twist0}), we obtain only
the modification of coproduct for Lorentz generators $M_{\mu \nu }$ 
\begin{align}
\Delta ^{\mathcal{F}}(P_{\mu })& =\Delta _{0}(P_{\mu }),  \label{cop0} \\
\Delta ^{\mathcal{F}}(M_{\mu \nu })& =\Delta _{0}(M_{\mu \nu })-\theta
^{\rho \sigma }[(\eta _{\rho \mu }P_{\nu }-\eta _{\rho \nu }\,P_{\mu
})\otimes P_{\sigma }  \label{cop1} \\
& \qquad \qquad \qquad \qquad \qquad +P_{\rho }\otimes (\eta _{\sigma \mu
}P_{\nu }-\eta _{\sigma \nu }P_{\mu })]\,.  \notag
\end{align}%
The coinverse as well as counit remain undeformed 
\begin{equation}
S(g)=-g,\qquad \epsilon (g)=0.  \label{se}
\end{equation}%
For twisted canonical space-time generated by twist (\ref{twist0}), the
multiplication $\star _{\mathcal{M}}$ (\ref{starr}) in the algebra $\mathcal{%
M}$ is defined by the twist factor $\mathcal{F}$ as follows 
\begin{equation}
e^{ipx}\star _{\mathcal{M}}e^{iqx}=m\circ \mathcal{F}^{-1}\vartriangleright
\lbrack e^{ipx}\otimes e^{iqx}]=m\circ (\overline{f}_{(1)}\rhd
e^{ipx}\otimes \overline{f}_{(2)}\rhd e^{iqy})=\mathcal{F}%
^{-1}(p,q)e^{ipx}e^{iqx},  \label{weyll}
\end{equation}%
where in general case we denote $\mathcal{F}^{-1}=\overline{f}_{(1)}\otimes 
\overline{f}_{(2)}$ and in canonical case $\mathcal{F}^{-1}(p,q)=e^{\,-\frac{%
i}{2}\,p\theta q},\,p\theta q\equiv p^{\mu }\theta _{\mu \nu }q^{\nu }$.

The algebra (\ref{kk}) can be extended consistently to the collection of
noncommutative space-time coordinates $(\widehat{x}_{\mu }^{(1)},\widehat{x}%
_{\mu }^{(2)}\ldots \widehat{x}_{\mu }^{(n)})$ in the following way \cite%
{wess}, \cite{pic} ($i,j=1,2,\ldots ,n$) 
\begin{equation}
\lbrack \widehat{x}_{\mu }^{(i)}\,,\widehat{x}_{\mu }^{(j)}]=i\theta _{\mu
\nu }\,.
\end{equation}%
If we introduce the corresponding Weyl map for bilocal product $e^{ip%
\widehat{x}}e^{iq\widehat{y}}$ $(\widehat{x}_{\mu }\equiv \widehat{x}_{\mu
}^{(1)},\widehat{y}_{\mu }\equiv \widehat{x}_{\mu }^{(2)})$ it can be shown
that the formula (\ref{weyll}) can be extended as follows%
\begin{equation}
e^{ipx}\star _{\mathcal{M}}e^{iqy}=m\circ \mathcal{F}^{-1}\vartriangleright
\lbrack e^{ipx}\otimes e^{iqy}]=\mathcal{F}^{-1}(p,q)e^{ipx}e^{iqy}\;.
\label{aas}
\end{equation}

The twisted canonical Poincare algebra is the triangular Hopf algebra, with
universal R matrix satisfying the relation 
\begin{equation}
\Delta _{21}=\mathcal{R}\Delta \mathcal{R}^{-1},  \label{brr}
\end{equation}%
and given by the formula 
\begin{equation}
\mathcal{R}=\mathcal{F}_{21}\mathcal{F}^{-1}=\mathcal{F}^{-2},\qquad 
\mathcal{R}_{21}=\mathcal{R}^{-1}=\mathcal{F}^{2},  \label{rfr}
\end{equation}%
where in (\ref{rfr}) we used the relation $\mathcal{F}_{21}=\mathcal{F}^{-1}$%
. The formulae (\ref{rfr}) describe the universal $\mathcal{R}$-matrix for
triangular Hopf algebras.

\section{Deformed quantum fields and braided tensor product}

If we construct deformed QFT we deal with the following three algebras: the
algebra $\mathcal{M}$ of functions $f\in \mathcal{M}$ on noncommutative
space time, the algebra $\mathcal{H}$ of deformed field oscillators $h\in 
\mathcal{H}$ and the algebra $\Phi $ of deformed fields $\phi \in \Phi .$The
deformed relativistic free quantum fields $\phi $\ can be described as the
infinite sum (in fact continuous integral) of the braided tensor products of
the noncommutative plane waves $e^{ip\widehat{x}}$ from $\mathcal{M}$ and
the elements $A(p)$ from the algebra of \ field oscillators $\mathcal{H}$%
\begin{equation}
{\phi }\in \mathcal{M}\underline{\otimes }\mathcal{H},  \label{gf}
\end{equation}%
where underlining of the tensor product described its braided nature.

The algebra $\mathcal{M}$($e^{ip\widehat{x}},\cdot $) of basic functions on
noncommutative space-time $\widehat{x}=(\widehat{x}_{i},\widehat{x}_{0})$ we
shall further represent izomorphically by the star product $\star _{\mathcal{%
M}}$ of commutative functions $e^{ipx}$ providing by means of the Weyl map $%
\mathcal{M}$($e^{ipx},\star _{\mathcal{M}}$)%
\begin{align}
& e^{ip\widehat{x}}e^{iq\widehat{x}}=m(e^{ip\widehat{x}}\otimes e^{iq%
\widehat{x}})  \notag \\
& \qquad \qquad \overset{Weyl\ map}{\longrightarrow }m_{\mathcal{M}%
}(e^{ipx}\otimes e^{iqx})=e^{ipx}\star _{\mathcal{M}}e^{iqx},  \label{starr}
\end{align}%
where $x=(x_{i},x_{0})$ describes standard Minkowski space-time coordinates.
We add that for the consideration of bilocal products ${\phi }(\widehat{{x}})%
{\phi }(\widehat{{y}})$ we should introduce as well the Weyl map $e^{ip%
\widehat{x}}e^{iq\widehat{y}}\longrightarrow e^{ipx}\star _{\mathcal{M}%
}e^{iqy}$ generalizing the formula (\ref{starr}) (see also (\ref{weyll}) and
(\ref{aas})). Effectively in formula (\ref{gf}) we shall describe $\mathcal{M%
}$ by the commutative functions on classical Minkowski space with the
modified star-multiplication law $m\rightarrow m_{\mathcal{M}}$.

The Poincare algebra as well as its Casimirs are not modified by the twist
deformation, i.e. if we consider only twisted Poincare symmetries they do
not change the mass-shell which is provided by the mass Casimir, or
equivalently the relativistic energy-momentum dispersion relation $\omega (%
\overrightarrow{p})=\sqrt{\overrightarrow{p}^{2}+m^{2}\text{ }}$remain valid
after twisting. In accordance with (\ref{gf}), in twisted field theory one
can use the standard Fourier decomposition of free scalar quantum fields
with noncommutative Fourier exponentials 
\begin{align}
{\phi }(\widehat{{x}})& =\frac{1}{(2\pi )^{4}}\int d^{4}p\delta
(p^{2}-m^{2})\;\mathrm{e}^{ip\widehat{{x}}}\otimes A(p)  \label{ffmm} \\
& =\frac{1}{(2\pi )^{3}}\int \frac{d^{3}\vec{p}}{2\omega (\vec{p})}\;\left(
\;\mathrm{e}^{ip\widehat{{x}}}\otimes a^{\dag }(\vec{p})+\mathrm{e}^{-ip%
\widehat{{x}}}\otimes a(\vec{p})\right) _{p_{0}=\omega (\vec{p})}\;,  \notag
\end{align}%
where $a(\vec{p})=A(\vec{p},\omega (\vec{p})),(a^{\dag }(\vec{p})=A(-\vec{p}%
,-\omega (\vec{p})))$ are the deformed annihilation (creation) field
oscillator.

\bigskip The algebra of deformed field oscillators $\mathcal{H(}A(p);\cdot )$
with standard multiplication in $\mathcal{H}$ ($m(A(p)\otimes A(q))=A(p)A(q)$%
) in noncommutative case will be described in accordance with the
quantum-deformed covariance condition by the braided commutation relations
of oscillators algebra with suitable multiplication in $\mathcal{H}$.

If we use the Weyl map (\ref{starr}) the elements of algebra $\mathcal{M}$
are replaced by classical functions (e.g. $e^{ip\widehat{x}}\rightarrow
e^{ipx}$). Further using the isomorphism $%
\mathbb{C}
\otimes \mathcal{H}\simeq \mathcal{H}$ one can introduce field $\phi (%
\widehat{x})\overset{W}{\simeq }\varphi (x)$ ($W$ denotes Weyl map) having
well-known standard form

\begin{align}
{\varphi }({x})& =\frac{1}{(2\pi )^{4}}\int d^{4}p\delta (p^{2}-m^{2})\;%
\mathrm{e}^{ip{x}}A(p)  \label{f2} \\
& =\frac{1}{(2\pi )^{3}}\int \frac{d^{3}\vec{p}}{2\omega (\vec{p})}\;\left(
\;\mathrm{e}^{ip{x}}a^{\dag }(\vec{p})+\mathrm{e}^{-ip{x}}a(\vec{p})\right)
_{p_{0}=\omega (\vec{p})}\;,  \notag
\end{align}%
where ${\varphi }({x})\in \mathcal{H}$, and we point out that oscillators $%
A(p)$ will satisfy the deformed binary relations.

The tensor structure of the deformed quantum field (see (\ref{gf}) and (\ref%
{ffmm})) should be consistent with the actions of the space-time symmetry
generator $g$. On the deformed symmetry algebra modules $\mathcal{M}$ and $%
\mathcal{H}$ the actions of $g$ is defined by the coproduct $\Delta
(g)=g_{(1)}\otimes g_{(2)}$ as follows%
\begin{eqnarray}
g\blacktriangleright a &=&ad_{g}a=\sum g_{(1)}aS(g_{(2)}).\qquad a\in 
\mathcal{H},  \label{adad} \\
g\rhd f &=&ad_{g}f=\widehat{D}(g)f.\qquad f\in \mathcal{M},  \label{adad2}
\end{eqnarray}%
where $\widehat{D}(g)$ describes the differential realization of $g$ on
noncommutative functions $f\in \mathcal{M}$.

If we wish to define the actions $\rhd $ and $\blacktriangleright $ on the
tensor product $\mathcal{M}\otimes \mathcal{H}$ we postulate in consistency
with the symmetry properties of no-deformation classical limit the trivial
actions on "wrong" part of the tensor product (\ref{gf})

\begin{eqnarray}
g\blacktriangleright f &=&\epsilon (g)f, \\
g\rhd a &=&\epsilon (g)a,
\end{eqnarray}%
where $\epsilon (1)=1$ and $\epsilon (g)=0$\ for any Poincare generators. We
see that modules$\mathcal{M}$ under action $\blacktriangleright $\ and the
module $\mathcal{H}$ under the action $\rhd $\ behaves as number (scalar
spectator). In particular e.g. if we choose the product of basic modules $%
e^{ip\widehat{x}}\underline{\otimes }a(\overrightarrow{p})$ one gets for
example%
\begin{eqnarray}
g\blacktriangleright (e^{ip\widehat{x}}\underline{\otimes }a(\overrightarrow{%
p})) &=&(ad\otimes ad)\Delta ^{\mathcal{F}}(g)(e^{ip\widehat{x}}\underline{%
\otimes }a(\overrightarrow{p}))  \label{e1} \\
&=&ad_{g_{(1)}}e^{ip\widehat{x}}\underline{\otimes }ad_{g_{(2)}}a(%
\overrightarrow{p})=(g_{(1)}\blacktriangleright e^{ip\widehat{x}})\underline{%
\otimes }(g_{(2)}\blacktriangleright a(\overrightarrow{p})).  \notag
\end{eqnarray}%
Because the general coproduct can be written as ($\Delta ^{(0)}\mathcal{(}g%
\mathcal{)}=g\otimes 1+1\otimes g$; see also (\ref{exx}))

\begin{equation}
\Delta ^{\mathcal{F}}\mathcal{(}g\mathcal{)}=\Delta ^{(0)}\mathcal{(}g%
\mathcal{)+}\text{ terms not containing components }1\otimes g\text{ and }%
g\otimes 1\text{,}  \label{esx}
\end{equation}%
the only term contributing to $g_{(1)}\blacktriangleright e^{ip\widehat{x}}$
comes from $1\otimes g$, and one gets

\begin{equation}
g\blacktriangleright (e^{ip\widehat{x}}\underline{\otimes }a(\overrightarrow{%
p}))=(1\blacktriangleright e^{ip\widehat{x}})\underline{\otimes }%
(g\blacktriangleright a(\overrightarrow{p}))=e^{ip\widehat{x}}\underline{%
\otimes }(g\blacktriangleright a(\overrightarrow{p})).  \label{aa1}
\end{equation}%
The relation (\ref{aa1}) has the correct classical limit when the factor $%
e^{ip\widehat{x}}$ becomes a classical function $e^{ipx}$. Similarly

\begin{equation}
g\rhd (e^{ip\widehat{x}}\underline{\otimes }a(\overrightarrow{p}))=(g\rhd
e^{ip\widehat{x}})\underline{\otimes }a(\overrightarrow{p}).  \label{aa2}
\end{equation}

In the consideration of deformed symmetry properties we shall need only the
actions (\ref{aa1}), (\ref{aa2}) of the generators $g$.

In this paper we consider the extension of the usual framework by
introducing the algebra of fields $\phi $ with braided multiplication rule.
We should multiply the noncommutative fields (\ref{gf}) in a way taking into
account the braid $\Psi _{\mathcal{M},\mathcal{H}}$ introduced as follows 
\begin{align}
& \phi (\widehat{x})\bullet \phi (\widehat{y})=m_{\mathcal{M}\underline{%
\otimes }\mathcal{H}}\phi (\widehat{x})\otimes \phi (\widehat{y})
\label{gmn} \\
& \qquad \qquad \quad =(m\otimes m)\circ (id\otimes \Psi _{\mathcal{M},%
\mathcal{H}}\otimes id)\,\phi (\widehat{x})\otimes \phi (\widehat{y}). 
\notag
\end{align}%
Braid factor $\Psi _{\mathcal{M},\mathcal{H}}$ describes effectively the
noncommutativity of factors $A(p)$ and $e^{iq\widehat{y}}$ in the product of
field operators in accordance with general formula for covariant braiding
acting on two modules of quasitriangular Hopf algebra. It is defined by
universal matrix $\mathcal{R=R}_{1}\otimes \mathcal{R}_{2}$ ($f\in \mathcal{M%
},h\in \mathcal{H}$); see e.g. \cite{sm}%
\begin{equation}
\Psi _{\mathcal{M},\mathcal{H}}(h\underline{\otimes }f)=(\mathcal{R}_{2}\rhd
f)\underline{\otimes }(\mathcal{R}_{1}\blacktriangleright h).  \label{bn}
\end{equation}%
We see that if wespecify in $\mathcal{M\underline{\otimes }H}$ the braid
factor $\mathcal{R}_{21}$, the generators $g$ (we recall that $\mathcal{R}$
is defined in terms of Poincare algebra generators $g$) act nontrivially on
both legs of the tensor product. In fact before applying Weyl map to the
product (\ref{gmn}) we should firstly use braid factor $\Psi _{\mathcal{M},%
\mathcal{H}}$ in order to introduce ordered elements of $\mathcal{M}$ and $%
\mathcal{H}$, with field oscillators on the right and noncommutative Fourier
exponentials on the left.

The covariant braid factor (\ref{bn}) will be necessary in order to
introduce the multiplication in the covariant algebra of deformed quantum
fields $\Phi (\phi ,m_{\mathcal{M}\underline{\otimes }\mathcal{H}})$. The
commutative (undeformed) limit of the relation (\ref{gmn}) is obtained by
putting $\Psi _{\mathcal{M},\mathcal{H}}=\tau $, where $\tau $ is the flip
operator describing no deformation limit of $\mathcal{R}_{21}$. In standard
Poincare symmetry case one assumes that%
\begin{equation}
\lbrack e^{ipx},A(q)]=(1-\tau )e^{ipx}A(q)=0.
\end{equation}%
i.e. the standard free quantum fields can be described by unbraided tensor
products ($m_{\mathcal{M}\underline{\otimes }\mathcal{H}}\longrightarrow m_{%
\mathcal{M}\otimes \mathcal{H}})$ which provides the ordering 
\begin{equation}
m_{\mathcal{M\underline{\otimes }H}}[(e^{ipx}\underline{\otimes }%
A(p))\otimes (e^{iqy}\underline{\otimes }A(q))]=e^{ipx}e^{iqy}\underline{%
\otimes }A(p)A(q)\simeq e^{i(px+qy)}A(p)A(q).
\end{equation}

In this paper we describe the algebra of deformed quantum free fields with
noncommutative space-time arguments endowed with nontrivial braided
multiplication (\ref{gmn}) consistent with twisted Poincare covariance. The
deformed covariant field quantization will be described by the $c$-number
braided field commutator (see Sect. 5). We shall perform the explicit
calculations for the case of quantum deformation described by the canonical
twist (\ref{twist0}). Firstly in following Section we shall describe the
covariance conditions satisfied by twist-deformed quantum fields.

\section{ The twisted covariance of single quantum field}

Let us consider firstly the Poincare covariance of the standard (undeformed)
free quantum field ${\varphi }({x})$, given by (\ref{f2}) with algebra $%
\mathcal{H}$ describing standard field oscillators algebra. The classical
Poincare covariance is given by the known formula%
\begin{equation}
U(\Lambda ,a)\varphi (x)U^{-1}(\Lambda ,a)=\varphi (\Lambda x+a),
\label{unitary}
\end{equation}%
where $U(\Lambda ,a)=\exp (ia^{\mu }P_{\mu }+i\omega ^{\mu \nu }M_{\mu \nu
}) $ describes the unitary representation of Poincare group in Hilbert-Fock
space, generated from the Poincare-invariant vacuum by the products of field
oscillators. For infinitesimal $a_{\mu }$ and $\omega _{\mu \nu }$ ($\Lambda
_{\ \nu }^{\mu }=\delta _{\ \nu }^{\mu }+\omega _{\ \nu }^{\mu }$) from
relation (\ref{unitary}) follows that there are two ways of expressing the
action of Poincare generators on quantum free field which give the same
result%
\begin{align}
P_{\mu }\blacktriangleright \varphi & :=ad_{P_{\mu }}^{(0)}\varphi =[P_{\mu
},\varphi ]=-D^{(0)}(P_{\mu })\varphi =:-P_{\mu }\rhd \varphi ,\qquad \qquad
\label{symme} \\
M_{\mu \nu }\blacktriangleright \varphi & :=ad_{M_{\mu \nu }}^{(0)}\varphi
=[M_{\mu \nu },\varphi ]=-D^{(0)}(M_{\mu \nu })\varphi =:-M_{\mu \nu }\rhd
\varphi .  \label{symme1}
\end{align}%
We see that the relations (\ref{symme})-(\ref{symme1}) can be written as

\begin{equation}
g^{(0)}\blacktriangleright \varphi =-g^{(0)}\rhd \varphi ,
\end{equation}%
where $g^{(0)}=$denotes the undeformed Poincare algebra generators provide
the generalized Heisenberg equation on the classical Poincare group manifold.

In undeformed quantum free field case the generators $P_{\mu },\ M_{\mu \nu
} $ are the bilinear functionals of the oscillators $a(p),a^{\dagger }(p)$
and on rhs of (\ref{symme})-(\ref{symme1}) we have the known relativistic
differential\ realizations $D^{(0)}(g_{A})$ $\ (g_{A}=(P_{\mu },\ M_{\mu \nu
}))$ of standard Poincare-Hopf algebra on the scalar fields with Minkowski
space-time arguments%
\begin{equation}
D^{(0)}(P_{\mu })=-i\partial _{\mu },\qquad \qquad D^{(0)}(M_{\mu \nu
})=-i(x_{\mu }\partial _{\nu }-x_{\nu }\partial _{\mu }).  \label{edif}
\end{equation}%
The generalized Heisenberg eq. (\ref{symme})-(\ref{symme1}) describe in
infinitesimal form the standard global Poincare covariance of quantum free
fields, given by relation (\ref{unitary}).

In noncommutative framework the description should be modified in accordance
with formula (\ref{gf}) and corresponding Hopf-algebraic language. The
deformed counterpart of the equations (\ref{symme}) and (\ref{symme1})
should be derived with the use of the formulae (\ref{e1})-(\ref{aa1}). If we
observe that the twisted coproduct is given by the formula (\ref{exx}), due
to the vanishing values of counit $\epsilon (g)$ only $\Delta ^{(0)}\mathcal{%
(}g\mathcal{)}$ in (\ref{exx}) will contribute to the adjoint action $%
g\blacktriangleright \phi $ in accordance with (\ref{esx}). We obtain
therefore in deformed case the following deformed generalized Heisenberg
equations

\begin{equation}
g\blacktriangleright \phi =S(g)\rhd \phi ,\qquad g=(P_{\mu },M_{\mu \nu })
\label{nw}
\end{equation}%
or more explicitly 
\begin{eqnarray}
P_{\mu }\blacktriangleright \phi &=&(ad_{P_{\mu (1)}}\otimes ad_{P_{\mu
(2)}})\phi =(1\otimes ad_{P_{\mu }})\phi =(D[S(P_{\mu })]\otimes 1)\phi
=S(P_{\mu })\rhd \phi ,\qquad  \label{dddf} \\
M_{\mu \nu }\blacktriangleright \phi &=&(ad_{M_{\mu \nu (1)}}\otimes
ad_{M_{\mu \nu (2)}})\phi =(1\otimes ad_{M_{\mu \nu }})\phi =(D[S(M_{\mu \nu
})]\otimes 1)\phi =S(M_{\mu \nu })\rhd \phi ,  \label{dddf2}
\end{eqnarray}%
where the adjoint action of deformed symmetry generators $g$ in $\mathcal{H}$
are given by (\ref{adad}) and $\widehat{D}(g)$ is the suitably deformed
differential realization on $\mathcal{M}$ of standard Poincare algebra,
consistent with deformed coalgebra structure. In particular after Weyl map $%
f(\widehat{x})\overset{W}{\rightarrow }f(x),$ $g(\widehat{x})\overset{W}{%
\rightarrow }g(x),$ $\widehat{D}(g)\overset{W}{\rightarrow }D(g)$ we should
have the following form of modified Leibnitz rule

\begin{equation}
D(g)(f(x)\star _{\mathcal{M}}g(x))=m_{\mathcal{M}}\circ \lbrack (D\otimes
D)\Delta ^{\mathcal{F}}(g)\circ f(x)\otimes g(x)].  \label{xex}
\end{equation}

In the case of canonical deformation, as follows from (\ref{cop0}), we
obtain undeformed covariance relation (\ref{symme}) for the momentum
generators, because from (\ref{adad}) follows that%
\begin{equation}
(1\otimes ad_{P_{\mu }})\phi =(1\otimes ad_{P_{\mu }}^{(0)})\phi ,
\label{zzz}
\end{equation}%
and $D(P_{\mu })=D^{(0)}(P_{\mu })$. For the Lorentz algebra generators $%
M_{\mu \nu }$ it follows from (\ref{cop1}), (\ref{se}) and (\ref{adad})\ that%
\begin{equation}
(1\otimes ad_{M_{\mu \nu }})\phi =[1\otimes M_{\mu \nu },\phi ]+\theta
_{\lbrack \nu }^{\quad \alpha }(1\otimes P_{\alpha })\phi (1\otimes P_{\mu
]})+\theta _{\lbrack \mu }^{\quad \alpha }(1\otimes P_{\nu ]})\phi (1\otimes
P_{\alpha }).  \label{mc2}
\end{equation}

In order to obtain in relation (\ref{dddf2}) the suitably deformed
differential realization of Lorentz algebra generators $D(M_{\mu \nu })$ one
can use or differential realization on noncommutative Minkowski space or
after performing the Weyl map we consider the deformation $D(M_{\mu \nu })$
of $D^{(0)}(M_{\mu \nu })$ (see (\ref{edif})) on standard Minkowski space
extended to relativistic phase space. We shall use below the second
possibility. Let us observe that the noncommutative space-time coordinates (%
\ref{kk}) can be described by the relativistic nondeformed quantum phase
space variables ($x_{\mu },p_{\mu }$) as follows%
\begin{equation}
\widehat{x}_{\mu }=x_{\mu }+\theta _{\mu \nu }p^{\nu },  \label{change}
\end{equation}%
where $[x_{\mu },p_{\nu }]=i\eta _{\mu \nu }(\eta _{\mu \nu }=$diag$%
(-1,1,1,1))$, $[x_{\mu },x_{\nu }]=[p_{\mu },p_{\nu }]=0.$ Using relation (%
\ref{change}) and the standard Lorentz transformations of the variables $%
x_{\mu },p_{\nu },$ the modified Lorentz transformation of canonically
deformed space-time variables $\widehat{x}_{\mu }$ can be expressed as%
\begin{equation}
\widehat{x}_{\mu }^{\prime }=\Lambda _{\mu }^{\ \nu }x_{\nu }+\theta _{\mu
\nu }\Lambda _{\rho }^{\ \nu }p^{\rho }=\Lambda _{\mu }^{\ \nu }\widehat{x}%
_{\nu }+\widehat{\xi }^{\nu }\,,
\end{equation}%
where $\widehat{\xi }^{\nu }=(\theta _{\mu \rho }\Lambda _{\ \nu }^{\rho
}-\Lambda _{\mu }^{\ \rho }\theta _{\rho \nu })p^{\nu }$ is the
momentum-dependent translation. Using infinitesimal Lorentz transformation $%
(\Lambda =\delta +\omega )$ we get for the noncommutative space-time
coordinates the following infinitesimal Lorentz transformations%
\begin{equation}
\widehat{x}_{\mu }^{\prime }=\widehat{x}_{\mu }+\omega _{\mu }^{\ \nu }%
\widehat{x}_{\nu }+\widehat{\zeta }_{\mu },  \label{xxx}
\end{equation}%
where $\widehat{\zeta }_{\mu }=(\omega ^{\rho \nu }\theta _{\mu \rho
}^{\quad }-\omega _{\mu }^{\ \rho }\theta _{\rho }^{\ \nu })p_{\nu }$. If we
perform the Weyl map $\widehat{x}_{\mu }\overset{W}{\rightarrow }x_{\nu }$
and observe from (\ref{change}) that $[\widehat{x}_{\mu },p_{\nu }]=i\eta
_{\mu \nu }\overset{W}{\rightarrow }[x_{\mu },p_{\nu }]=i\eta _{\mu \nu }$,
the canonically conjugated momenta $p_{\mu }$ can be represented by the
derivatives ($p_{\mu }\rightarrow -i\partial _{\mu })$ and formula (\ref%
{ffmm}) is replaced by (\ref{f2}). Then the infinitesimal deformed Lorentz
transformation, as follows from (\ref{xxx}), imply the following change of
the field operator (\ref{f2}) 
\begin{equation}
i\omega ^{\mu \nu }D(M_{\mu \nu })\varphi =[\omega ^{\alpha \beta }(\delta
_{\alpha }^{\mu }x_{\beta }-\delta _{\beta }^{\mu }x_{\alpha })+\widehat{%
\zeta }^{\mu }]\partial _{\mu }\varphi ,
\end{equation}%
and we obtain that 
\begin{equation}
D(M_{\mu \nu })=-i(x_{\mu }\partial _{\nu }-x_{\nu }\partial _{\mu
})-[\theta _{\nu }^{\ \rho }\partial _{\rho }\partial _{\mu }+\theta _{\,\mu
}^{\rho }\partial _{\nu }\partial _{\rho }].  \label{llor}
\end{equation}%
We recall here that the formula (\ref{llor}) is known \cite{wess},\cite{kt}.
Expressing the relation (\ref{llor}) for deformed differential realization
as the relation between classical Poincare algebra generators we get the
deformed Lorentz generators as given by nondeformed Poincare generators as
follows (we recall that $P_{\mu }=P_{\mu }^{(0)}$) 
\begin{equation}
M_{\mu \nu }=M_{\mu \nu }^{(0)}+\theta _{\nu }^{\ \rho }P_{\rho }P_{\mu
}+\theta _{\;\mu }^{\rho }P_{\nu }P_{\rho }.
\end{equation}

The covariance relations (\ref{dddf}) describe the equality of the action on
the field (\ref{ffmm}) of the Poincare algebra symmetry generators $g$ in
classical (differential) space-time and quantum-mechanical (adjoint) field
oscillators realizations. For arbitrary twist deformation the quantum
adjoint action (\ref{adad}) and differential field realizations are
determined by the coalgebra relations of quantum Poincare symmetry (see (\ref%
{ppoc})). Several explicit examples of differential realizations $D(g)$
satisfying (\ref{dddf}), (\ref{xex}) has been presented in literature, but
the formulae providing $D(g)$ for arbitrary twist were not given\footnote{%
See however \cite{bp} where the generalization of formula (\ref{change}) for
arbitrary twist is proposed}.

\section{Braided multiplication of fields and covariant field commutator}

\subsection{\protect\bigskip The covariance of binary product of
noncommutative free quantum fields}

The bilocal $\star _{\mathcal{M}}$-multiplication for canonically deformed
Poincare symmetry is given by the formula (\ref{aas}). In the algebra of
oscillators $\mathcal{H}(A(p),\cdot )$ we use the standard multiplication
rule.

We shall introduce the braid factor $\Psi _{\mathcal{M},\mathcal{H}}$ in the
physical basis of $\mathcal{M}$ and $\mathcal{H}$ in accordance with (\ref%
{bn})%
\begin{equation}
\Psi _{\mathcal{M},\mathcal{H}}(A(p)\underline{\otimes }e^{iq\widehat{y}})=%
\mathcal{R}_{(2)}\rhd e^{iq\widehat{y}}\underline{\otimes }\mathcal{R}%
_{(1)}\blacktriangleright A(p).  \label{bgg}
\end{equation}%
For twist-deformed theory the multiplication prescription (\ref{gmn}) is
determined if \ we know the twist factor $\mathcal{F}$ and the braid $\Psi _{%
\mathcal{M},\mathcal{H}}$ given by (\ref{bgg}). The explicit form of the
product (\ref{gmn}) of fields on canonical noncommutative space-time has
therefore a form 
\begin{align}
\phi (\widehat{x})\bullet \phi (\widehat{y})& =m_{\mathcal{M}\underline{%
\otimes }\mathcal{H}}[\phi (x)\otimes \phi (y)]  \label{aaaa} \\
& =\frac{1}{(2\pi )^{8}}\int d^{4}p\int d^{4}q\delta (p^{2}-m^{2})\delta
(p^{2}-m^{2})  \notag \\
& \qquad \qquad \qquad e^{ip\widehat{x}}(\mathcal{R}_{(2)}\triangleright
e^{iq\widehat{y}})\underline{\otimes }(\mathcal{R}_{(1)}\blacktriangleright
A(p))A(q),  \notag \\
& \overset{W}{\simeq }\frac{1}{(2\pi )^{8}}\int d^{4}p\int d^{4}q\delta
(p^{2}-m^{2})\delta (p^{2}-m^{2})  \notag \\
& \qquad \qquad \qquad (\overline{f}_{(1)}\triangleright e^{ipx})(\overline{f%
}_{(2)}\mathcal{R}_{(2)}\triangleright e^{iqy})(\mathcal{R}%
_{(1)}\blacktriangleright A(p))A(q),  \notag
\end{align}%
where we recall that $\mathcal{F}^{-1}=\overline{f}_{(1)}\otimes \overline{f}%
_{(2)}$\ and by the notation $\overset{W}{\simeq }$ we denote the Weyl
homomorphism in $\mathcal{M}$ with the $\star _{\mathcal{M}}$-product of
classical Fourier exponentials \ representing the product $e^{ip\widehat{x}%
}e^{iq\widehat{y}}$. In the actions on the classical plane waves which are
present in last formula (\ref{aaaa}) one uses the differential realization $%
D(g_{A})$ (for deformation by canonical twist see (\ref{llor})), and on the
field oscillators the Poincare generators act by the quantum adjoint action
(see (\ref{adad}), (\ref{zzz}-\ref{mc2})). Before explicit calculation of
braided commutator (\ref{fgf}) we shall show that it is twist-covariant.
Using formula (\ref{vvoc}) for the algebra $\Phi (\phi ,m_{\mathcal{M}%
\underline{\otimes }\mathcal{H}})$ of deformed free quantum fields we choose%
\footnote{%
Using the covariance condition given by (\ref{nw}) one can choose
alternatively in (\ref{f1})-(\ref{f3}) the actions of symmetry generators on 
$\mathcal{H}$ by the actions on $\mathcal{M}$.} 
\begin{align}
g\blacktriangleright (\phi (\widehat{{x}})\bullet \phi (\widehat{{y}}))& =m_{%
\mathcal{M}\underline{\otimes }\mathcal{H}}[\Delta (g)\blacktriangleright
\phi (\widehat{x})\otimes \phi (\widehat{{y}})]  \label{f1} \\
& =(g_{(1)}\blacktriangleright \phi (\widehat{x}))\bullet
(g_{(2)}\blacktriangleright \phi (\widehat{{y}})),  \notag
\end{align}%
where the quantum action (\ref{adad}) on the algebra $\mathcal{H}$ is used.
Further ($\mathcal{R}_{21}\mathcal{=R}_{(2)}\otimes \mathcal{R}_{(1)}$) 
\begin{align}
g\blacktriangleright (\mathcal{R}_{21}\blacktriangleright \lbrack \phi (%
\widehat{{y}})\bullet \phi (\widehat{x})])& =g\blacktriangleright \lbrack (%
\mathcal{R}_{(2)}\blacktriangleright \phi (\widehat{{y}}))\bullet (\mathcal{R%
}_{(1)}\blacktriangleright \phi (\widehat{x}))] \\
& =m_{\mathcal{M\underline{\mathcal{\otimes }}H}}(\Delta (g)\mathcal{R}%
_{21}\blacktriangleright \phi (\widehat{{y}})\otimes \phi (\widehat{x})). 
\notag
\end{align}%
The twisted covariance of braided commutator (\ref{fgf}) means that 
\begin{equation}
g\blacktriangleright \lbrack \phi (\widehat{x}),\phi (\widehat{{y}}%
)]_{\bullet }^{BR}\equiv (g_{(1)}\blacktriangleright \phi (\widehat{x}%
))\bullet (g_{(2)}\blacktriangleright \phi (\widehat{{y}}))-\mathcal{R}%
_{21}\blacktriangleright \lbrack (g_{(2)}\vartriangleright \phi (\widehat{{y}%
}))\bullet (g_{(1)}\blacktriangleright \phi (\widehat{x}))],  \label{f3}
\end{equation}%
what requires the choice (\ref{bgg}) of the braid factor defining $\bullet $%
-multiplication and the validity of the relation (see also (\ref{brr})) 
\begin{equation}
\Delta (g)\mathcal{R}_{21}-\mathcal{R}_{21}\Delta _{21}(g)=0\,.  \label{kkom}
\end{equation}%
Indeed, if we use the formulas (\ref{ppoc}) and (\ref{rfr}) one shows easily
that in twist-deformed theory the relation (\ref{kkom}) is valid\footnote{%
In fact the relation (\ref{kkom}) remains true and the formula (\ref{fgf})
for covariant braided commutator of deformed fields is correct for any
quasitriangular Hopf algebra with the universal $\mathcal{R}$-matrix.} 
\begin{equation}
\Delta (g)\mathcal{R}_{21}-\mathcal{R}_{21}\Delta _{21}(g)=\mathcal{F}%
[\Delta _{0}(g),\tau ]\mathcal{F}^{-1}=0\,.
\end{equation}

\subsection{\protect\bigskip Calculation of covariant braided field
commutator: braided field oscillators algebra and braided locality.}

By using (\ref{aaaa}) and (\ref{fgf}) we shall calculate explicitly the
braided commutator. We get 
\begin{align}
\lbrack \phi (\widehat{x}),\phi (\widehat{y})]_{\bullet }^{BR}& =\phi (%
\widehat{x})\bullet \phi (\widehat{y})-\mathcal{R}_{21}\blacktriangleright
(\phi (\widehat{y})\bullet \phi (\widehat{x}))  \label{big} \\
& \overset{W}{\simeq }\frac{1}{(2\pi )^{8}}\int d^{4}p\int d^{4}q\delta
(p^{2}-m^{2})\delta (q^{2}-m^{2})  \notag \\
& \qquad \qquad \qquad \lbrack (\overline{f}_{(1)}\triangleright e^{ipx})(%
\overline{f}_{(2)}\mathcal{R}_{(2)}\triangleright e^{iqy})(\mathcal{R}%
_{(1)}\blacktriangleright A(p))A(q)  \notag \\
& \qquad \qquad \qquad \qquad -(\overline{f}_{(1)}\triangleright e^{iqy})(%
\overline{f}_{(2)}\mathcal{R}_{(2)}\triangleright e^{ipx})(\mathcal{R}_{(1)}%
\mathcal{R}_{(2)}\blacktriangleright A(q))(\mathcal{R}_{(1)}%
\blacktriangleright A(p)).  \notag
\end{align}

\bigskip Let us consider further the canonical deformation described by the
twist (\ref{twist0}). As follows from (\ref{rfr}) $\mathcal{R}_{21}$ depends
only on the fourmomentum generators and their actions follow from the
formulae%
\begin{equation}
P_{\mu }\rhd e^{ipx}=p_{\mu }e^{ipx}\,,\qquad P_{\mu }\blacktriangleright
A(p)=-p_{\mu }A(p).\,
\end{equation}%
In canonically deformed case the universal $\mathcal{R}$-matrix (\ref{rfr})
acts as follows%
\begin{eqnarray}
\mathcal{R}_{21}\rhd \lbrack e^{ipx}\otimes e^{iqy}] &=&e^{\,i\,\theta ^{\mu
\nu }\,p_{\mu }q_{\nu }}\,\,e^{ipx}\otimes e^{iqy},  \label{fun} \\
\mathcal{R}_{21}\blacktriangleright \lbrack A(p)\otimes A(q)]
&=&e^{\,i\,\theta ^{\mu \nu }\,p_{\mu }q_{\nu }}\,A(p)\otimes A(q)\,,  \notag
\end{eqnarray}%
and the braid $\Psi _{\mathcal{M},\mathcal{H}}$ has the explicit form

\begin{equation}
\Psi _{\mathcal{M},\mathcal{H}}[e^{ipx}\mathcal{\underline{\mathcal{\otimes }%
}}A(q)]=\mathcal{R}_{(2)}\rhd e^{ipx}\mathcal{\underline{\mathcal{\otimes }}R%
}_{(1)}\blacktriangleright A(q)=e^{\,-i\,\theta ^{\mu \nu }\,p_{\mu }q_{\nu
}}\,\,e^{ipx}\mathcal{\underline{\mathcal{\otimes }}}A(q).\,
\end{equation}

In order to obtain $c$-number braided field commutator one should be able to
factor out in (\ref{big}) the binary relations satisfied by the field
oscillators (the field oscillators algebra). If we use the formula (\ref{fun}%
) the required factorization in formula (\ref{big}) is achieved\ (we use in
last term of rhs of (\ref{ccom}) the short-hand notation described in
footnote 2)%
\begin{align}
\lbrack \phi (x),\phi (y)]_{\bullet }^{BR}& \overset{W}{\simeq }\frac{1}{%
(2\pi )^{8}}\int d^{4}p\int d^{4}q\delta (p^{2}-m^{2})\delta
(q^{2}-m^{2})e^{ipx}e^{iqy}  \label{ccom} \\
& \qquad \qquad \lbrack A(p)\star _{\mathcal{H}}A(q)-\mathcal{R}%
_{21}\blacktriangleright (A(q)\star _{\mathcal{H}}A(p))],  \notag
\end{align}%
provided that we introduce the following multiplication for the description
of binary products%
\begin{equation}
A(p)\star _{\mathcal{H}}A(q)=m\circ \mathcal{F}\blacktriangleright \lbrack
A(p)\otimes A(q)]=e^{\,\frac{i\,}{2}\theta ^{\mu \nu }\,p_{\mu }q_{\nu
}}A(p)A(q).  \label{nsm}
\end{equation}%
We add that the nonstandard multiplication $\star _{\mathcal{H}}$ is
different from $\star _{\mathcal{M}\text{ }}$($\mathcal{F}^{-1}$ in (\ref%
{weyll}) is replaced by $\mathcal{F}$) but it is known from the literature
and was used e.g. in \cite{wess}.

We recall that in undeformed theory the covariant formulation of the field
oscillator algebra is provided by the relation (see e.g. \cite{bs})

\begin{equation}
\delta (p^{2}-m^{2})\delta (q^{2}-m^{2})[A(p),A(q)]=\epsilon (p_{0})\delta
(p^{2}-m^{2})\delta ^{(4)}(p+q).  \label{gclas}
\end{equation}%
The following modification of relation (\ref{gclas}) describes the binary
relation for deformed field oscillators which due to the presence of braid
factor $\mathcal{R}_{21}$ is covariant under quantum symmetries and leads to 
$c$-number value of the braided commutator (\ref{ccom})%
\begin{equation}
\delta (p^{2}-m^{2})\delta (q^{2}-m^{2})[A(p)\star _{\mathcal{H}}A(q)-%
\mathcal{R}_{21}\blacktriangleright (A(q)\star _{\mathcal{H}}A(p))]=\epsilon
(p_{0})\delta (p^{2}-m^{2})\delta ^{(4)}(p+q).  \label{ggh}
\end{equation}

If we substitute (\ref{ggh}) into (\ref{ccom}) we obtain that%
\begin{equation}
\lbrack \phi (x),\phi (y)]_{\bullet }^{BR}\overset{W}{\simeq }\Delta
(x-y;m^{2}),  \label{loc}
\end{equation}%
with the braided commutator for canonically deformed free quantum fields
given by the known classical Pauli-Jordan function 
\begin{equation}
\Delta (x-y;m^{2})=\frac{-i}{(2\pi )^{3}}\int \frac{d^{3}p}{\omega (\vec{p})}%
\sin [\omega (\vec{p})(x_{0}-y_{0})]e^{i\vec{p}(\vec{x}-\vec{y})}\,.
\end{equation}

It should be noted that the choice of $\star _{\mathcal{M}}$(see (\ref{weyll}%
) for canonical case) and of the covariant braid $\Psi _{\mathcal{M},%
\mathcal{H}}$ (see (\ref{bn}) and (\ref{bgg})) is necessary for getting the
twist-covariant algebra of deformed field operators. The braid factor $%
\mathcal{R}_{21}$ which is an intertwiner in quantum quasitriangular
Poincare-Hopf algebra appears in our framework on three levels:

\begin{enumerate}
\item in the Weyl realization of the algebra $\mathcal{M}$ as expressing the
"braided commutativity" of the $\star _{\mathcal{M}}-$multiplication (see
e.g. \cite{alv};$\ f,h\in \mathcal{M}$) 
\begin{equation}
\lbrack f,h]_{\star _{\mathcal{M}}}^{BR}:=f\star _{\mathcal{M}}h-(\mathcal{R}%
_{(2)}\vartriangleright h)\star _{\mathcal{M}}(\mathcal{R}%
_{(1)}\vartriangleright f)=0\text{.}\qquad  \label{brk}
\end{equation}

In particular putting $f=x_{\mu },$ $g=x_{\nu }$ in (\ref{brk}) we reproduce
the space-time noncommutativity corresponding to given $\mathcal{R}$. We add
that the relation (\ref{brk}) is covariant for any choice of $f$ and $h$
under the action of deformed Poincare generators $g=(P_{\mu },M_{\mu \nu })$ 
\begin{equation}
g\vartriangleright \lbrack f,h]_{\star _{\mathcal{M}}}^{BR}=0\text{.}\qquad
\end{equation}

\item in the algebra (\ref{ggh}) of quantized field oscillators which is
covariant under the action $\blacktriangleright $ of the symmetry generators 
$g$.

\item in the algebra of deformed free quantum field $\phi $ (see (\ref{fgf}%
)) and the multiplication $m_{\mathcal{M}\otimes \mathcal{H}},$ what leads
to the covariance of braided field commutator (\ref{loc}).
\end{enumerate}

\bigskip

We point out that the nonstandard multiplication (\ref{nsm}) is selected by
the validity of braided $\bullet -$locality described by formula (\ref{loc}%
). If we look at the formula (\ref{ggh}) from the point of view of its
deformed Poincare covariance one can show that olny the braid factor $%
\mathcal{R}_{21}$ is required, but the choice of multiplication in $\mathcal{%
H}\ $is not determined.

\section{\protect\bigskip Final remarks}

\bigskip

The aim of this paper was to study the quantum Poincare covariance in the
framework of braided formulation of the theory of noncommutative quantum
free fields. We restricted our considerations to binary products of such
fields, but for twist-deformed noncommutative fields the extension of our
formalism to n-ary associative products is straightforward. It can be shown
(see e.g. \cite{maj}) that the associativity of braided products will follow
from the hexagon relation satisfied by braid $\mathcal{R}_{21}$.

We stress that we employ triple twisted covariance requirement, obtained the
braided form (\ref{fgf}) of \ deformed field commutator, the braided form (%
\ref{ggh}) of deformed oscillator algebra and braided product (\ref{gmn}) in
the algebra of deformed quantum free fields. We show that the derivation of
covariant braided $c$-number field commutator describing braided locality
requires the validity of the relation (\ref{ggh}) for deformed oscillators.
It should be stressed that separate elements of our construction were
present in previous papers (e.g. the Weyl map (\ref{starr}) is used in all
papers with the realization of noncommutative fields in standard Minkowski
space; see also \cite{dim},\cite{fi}) but all elements of our construction
occur together only in this paper.

Our explicit calculations have been given for the simplest case of canonical
twist deformation, with the additional numerical phase space factors in
momentum space characterizing the canonical deformation. If however the
twist factor depends as well on the Lorentz generators $M_{\mu \nu }$ (see
e.g. \cite{lw}), the formulae describing deformed fields are more
complicated. In such a case due to the realization (\ref{edif}) of Lorentz
generators after the Weyl map (\ref{starr}) the bidifferential operator
describing star product $\star _{\mathcal{M}}$ depends as well on the
space-time coordinate $x_{\mu }$. Analogously the adjoint action of the
braid factor $\mathcal{R}_{21}$ on oscillators in relations (\ref{ggh}) is
described effectively by the bidifferential operators in four-momentum space
acting on the product of two field oscillators which are the operator-valued
functions depending on the four-momenta $p_{\mu },q_{\mu }$ \footnote{%
See e.g. the derivation of nonlocal product of $\kappa $-deformed field
oscillators given in \cite{gen}}.

\bigskip Important question which should be considered in the future is the
application of braided-deformed free quantum fields for the description of
deformed interacting QFT. For that purpose one should use the deformed
version of the formulae expressing interacting quantum fields in terms of
free fields, within the perturbative framework. There are two ways of
approaching this problem:

\begin{enumerate}
\item One can deform the perturbative rules for Feynman diagrams, with
braid-deformed free Feynman propagators and suitably modified vertices. Such
perturbative description in case of canonical deformation however without
the use of braiding and quantum group symmetries was proposed firstly by
Filk \cite{fil}, and leads to nonlocal generalization of Dyson S-matrix
formula (the nonlocality is obtained if we replace after the Weyl map the
standard point-wise multiplication of fields by nonlocal $\star $%
-multiplication). At present it is a challenge to derive braid-deformed
counterpart of Dyson formula describing deformed perturbative S-matrix
expansion (for clue in this direction see e.g. \cite{sas},\cite{blom}).

\item Other way of defining deformed interacting QFT is to modify the
formulae expressing the interacting fields in terms of free asymptotic
fields (they define so-called Haag expansion \cite{haag}). The basic
dynamical tool for such approach is provided by perturbative solution of
deformed Yang-Feldmann equation.
\end{enumerate}

Finally let us observe that, the presented formalism with deformed braided
fields in principle can be applied to general quasitriangular quantum
deformation of free quantum fields. Such general deformations can be
described by the use of twist technique only in the Drinfeld category of
quasi-Hopf algebras, with nontrivial coassociator introducing nonassociative
product of three field operators (\cite{dfd}-\cite{maj},\cite{yz1}). 
\begin{equation}
\phi (\widehat{x})\bullet (\phi (\widehat{y})\bullet \phi (\widehat{z}%
))\equiv \Psi _{1(23)}(\phi (\widehat{x})\bullet \phi (\widehat{y}))\bullet
\phi (\widehat{z}).
\end{equation}%
In particular because the $\kappa $-deformation of Poincare symmetries is
not described by triangular Poincare-Hopf algebra, the description of $%
\kappa $-deformed free quantum fields covariant under $\kappa $-deformed
Poincare symmetries can be described by twist only with nontrivial
coassociator and requires the application of non-coassociative framework of
quasi-Hopf algebras (\cite{yz1}, \cite{yz2}). We should mention that
effective application of the approach presented here to the $\kappa $%
-deformed field-theoretic framework is under consideration.

\bigskip

\textbf{Acknowledgements: }

We would like to thank Paolo Aschieri and Gaetano Fiore for valuable
comments. The paper has been supported by Ministry of Science and Higher
Education by grant NN 202331139, and at final stage by NCN grant
2011/01/B/ST2/03354.

\end{document}